\documentclass{ws-ijmpcs}

\begin{document}

\markboth{Daniel C. Homan}
{Physical Properties of Jets in AGN}

%
\catchline{}{}{}{}{}
%

\title{Physical Properties of Jets in AGN}

\author{Daniel C. Homan}

\address{Department of Physics and Astronomy, Denison University\\
Granville, OH 43023,
USA
homand@denison.edu}

%

\maketitle

\begin{history}
\received{Day Month Year}
\revised{Day Month Year}
\end{history}

\begin{abstract}
I review constraints on the physical properties of AGN jets revealed
through Very Long Baseline Interferometry (VLBI) studies of the 
structure and time-evolution of parsec-scale jets, including 
recent results from the MOJAVE program.  In particular I focus on 
constraints available from very long {\em time} baseline studies which
probe a wide range of jet behavior over many outbursts.  Kinematic
studies of propagating jet features find an apparent speed distribution
that peaks around 10$c$ for blazars, with speeds up to 50$c$ observed.  These
observed speeds require Lorentz factors at least as large, implying that
parsec-scale Lorentz factors up to 10$-$20 are common for blazars with a
tail up to $\sim 50$.  Jet flows are still 
becoming organized on these scales as evidenced by the high incidence of 
non-radial motions and/or accelerations of jet features (including 
increases and decreases in apparent speed and direction).  Changes in 
Lorentz factors of propagating jet features
appear to play a significant role in the observed accelerations, and while
the connection between acceleration of jet features and the underlying flow 
is not clear, the pattern of observed accelerations suggest the 
flow may increase in speed near the base of the jet and decrease further 
out.  In some jets, ejections of new features span a range of ejection 
angles over many epochs, tracing out wider opening angles on parsec-scales 
than are apparent in single epoch observations.
\end{abstract}

\ccode{PACS numbers: 11.25.Hf, 123.1K}

\section{Introduction}	

A key goal of Very Long Baseline Interferometry (VLBI) studies of AGN 
jets is to address long standing questions about jet formation, collimation, 
and acceleration\cite{Meier01}\cdash\cite{MJ08}.  These studies also seek to understand the distribution of intrinsic 
jet power and speeds\cite{VC94,LM97,CLH07} and the connection between those properties and 
other high energy phenomena\cite{K09}\cdash\cite{MJL10}.
VLBI observations of AGN jets directly probe
the structure, polarization, and kinematics of extra-galactic radio
jets on parsec scales. Because jets are moving very close to the speed
of light and directed close to our line of sight, their emission
is highly Doppler boosted, and features moving within the jets exhibit
apparent superluminal motion: 

\begin{equation}
\beta_{obs} = \frac{\beta\sin\theta}{1-\beta\cos\theta}
\end{equation}

\vspace{0.15in}

\noindent where $\beta$ is the intrinsic speed of the moving feature and
$\theta$ in the angle its motion makes with our line of sight.  For 
a given $\beta$, this motion is a maximum when $\beta=\cos\theta$, 
in which case $\beta_{obs} = \beta\Gamma$ where $\Gamma = 1/\sqrt{1-\beta^2}$ 
is the Lorentz factor of the moving feature. 

An important question is the extent to which any single jet feature or `component' is 
characteristic of the jet flow.  It is now commonly believed that most jet features 
represent propagating oblique or transverse shocks in the flow\cite{MG85}\cdash\cite{HAA11}.
Consistent with this picture, a wide range of behavior is 
seen among components within a single jet, including stationary or 
quasi-stationary features\cite{H01}\cdash\cite{LCH09},
although it is important to note that 
Ref.~\refcite{LCH09} found the distribution of apparent speeds within individual 
jets to be significantly narrower than between different jets, suggesting 
that individual jets do have a characteristic flow speed. When multiple 
speeds are seen in a single jet, slower speeds may represent trailing shocks\cite{G05}, 
and therefore the best estimate of the flow speed may be the 
fastest component observed\cite{LCH09}.

The National Radio Astronomy Observatory's\footnote{The National Radio Astronomy 
Observatory is a facility of the National Science Foundation operated under 
cooperative agreement by Associated Universities, Inc.}  Very Long Baseline Array (VLBA) 
has been in continuous operation since 1994, allowing regular monitoring
of large samples of AGN radio jets, with some individual jets having regular monitoring
observations for 16 years and counting. The large sample sizes and very long {\em time} 
baselines allowed by the continuous operation of the VLBA has made it possible 
to study AGN jets in entirely new ways to address these questions. In the following
sections I will discuss the impact on our understanding of (1) the distribution
of apparent jet speeds and intrinsic Lorentz factors, (2) the acceleration and
collimation of jets, and (3) the morphology and opening angles of parsec-scale jets. 

\section{Long Time Baseline VLBI}

As described above, the VLBA has made it possible to study both large samples of 
AGN jets and to continuously monitor jets over long time intervals.  The longest 
running example of this kind of program is the 2cm Survey/MOJAVE program.  MOJAVE
stands for Monitoring Of Jets from Active galactic nuclei with VLBA Experiments\cite{LH05}\cdash\cite{LAA09},
and it is a continuation of 2cm Survey which started in 1994\cite{K98,Z02,K04,K05}.  The MOJAVE program
currently images the parsec scale structure and polarization of more than 300 AGN jets at 
15 GHz ($\lambda 2$ cm) on a regular basis. Many of the jets in the 135 source MOJAVE-I
complete, flux-density limited sample\cite{LAA09} have observations spanning more than 
15 years. The Gamma-ray AGN monitoring project at Boston University has focused on
monitoring a smaller number of Gamma-ray Blazars identified by EGRET and FERMI at
higher frequencies (22 and 43 GHz) with more closely spaced epochs\cite{J01,J05,MJL10}.  
While it is difficult
to track individual jet features over long periods of
time at these higher frequencies, several of the AGN in their program also have data spanning more than 15 years.  
At lower frequencies (2 and 8 GHz) the Radio Reference Frame Image Database 
(RRFID) has images spanning a similar period of time\cite{F96}; however, only a small 
window from 1994-1998 has been analyzed for kinematics\cite{PMF07}.

Figure 1 is an example of the kind of the kinematic data that can
be obtained by observing a parsec scale jet over long periods of time.  The
TeV Blazar 1222+216 has 15 GHz monitoring observations going back to 1996 as part of
the Brandeis University monitoring project\cite{H01,O04} and has been observed as part
of the 2cm Survey/MOJAVE program since 1999.  The MOJAVE program identifies several
moving components that can be tracked throughout much of this period\cite{LCH09}, and the motion of
component ``5'' is illustrated in the figure.  It has an apparent motion nearly
17$c$ and makes a distinct bend in its trajectory to the East, magnified by
projection to appear to be a $40^\circ$ change in the plane of the sky\cite{LIP}.  

\begin{figure}[pb]
\centerline{\psfig{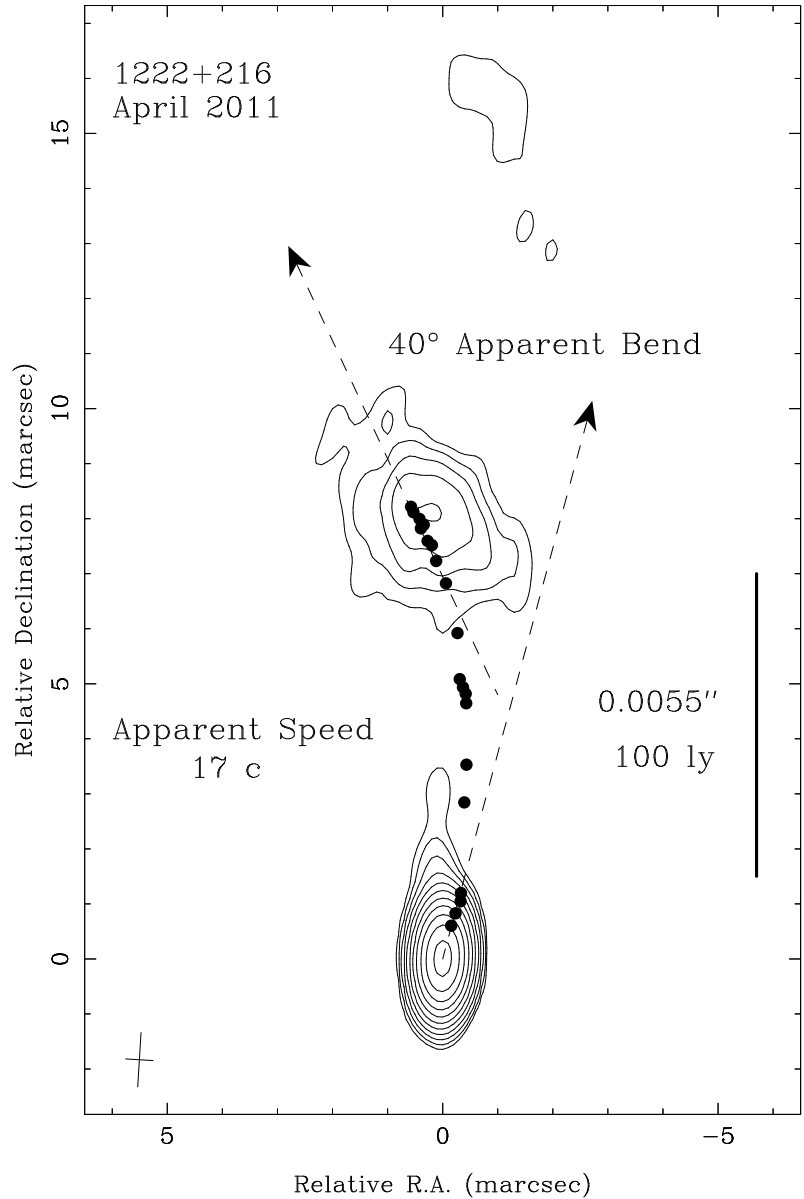}}
\vspace*{8pt}
\caption{Plot of 12 years of MOJAVE position data for jet component ``5'' in the TeV Blazar 1222$+$216 
at $z=0.434$.  The component has an apparent speed of nearly 17$c$, and shows a marked bending of 
its trajectory to the East.  The positions are superposed on a 15 
GHz MOJAVE VLBA image from April 2011 (http://www.physics.purdue.edu/astro/MOJAVE/).  Contours begin 
at 1.5 mJy/beam and increase in factor of two steps.\label{f1}}
\end{figure}

\subsection{Distribution of Apparent Speeds}

By studying the apparent speed distribution of a large, flux-density limited sample of AGN jets we probe the
underlying Lorentz factor distribution of the parent population\cite{VC94,LM97}.  This assumes that the
apparent speeds of moving jet features are good tracers of the underlying flow velocity, and
as discussed in the introduction, the range of apparent speeds seen in individual objects
suggests that we should use only the largest apparent speed observed in a given jet.  Thus it
is necessary to observe jets long enough to sample multiple components to increase our confidence
that we have seen a component characteristic of the flow speed.  In their analysis of the 135
source MOJAVE-I complete sample, Ref.~\refcite{LCH09} obtained motions for 127 AGN jets and found that the 
distribution of the fastest component in each jet peaked around 10$c$ with a tail extending up 
to 50$c$. Analysis of the effects of beaming on flux-density limited samples show that for 
a large sample, like the MOJAVE-I sample, the fastest observed speed in the sample is characteristic
of the maximum Lorentz factor in the parent population\cite{LM97}; thus, Ref.~\refcite{LCH09} 
concludes that the intrinsic Lorentz factor distribution is likely a power-law distribution with a 
tail extending up to $\Gamma \simeq 50$. It is interesting to note that a jet beamed directly at
us has a Doppler factor $\delta = (1+\beta)\Gamma$, so Doppler factors of up to $\sim 100$, although rare,
may be observed in the blazar population. 

Ref.~\refcite{J05} find a similar range of speeds in their sample of 15 blazars at $\sim3\times$ 
higher resolution, suggesting that the much larger MOJAVE sample is indeed sensitive to the fastest jet 
speeds emerging from the core region, despite the resolution difference.  Ref.~\refcite{J05} also 
performed an additional analysis of the apparent fading times of the jet components they followed 
in an attempt to obtain an independent estimate of the Doppler factor for individual components.  
The combination of Doppler factor and apparent speed for individual components allowed a direct 
estimate of the Lorentz factors for those components which fell in the range $\Gamma = 5 - 40$.  

The lower frequency Caltech-Jodrell Bank Flat-Spectrum (CJF) sample of speeds in 237 AGN at 5 GHz\cite{B08} 
and the RRFID kinematic analysis at 8 GHz of 77 sources\cite{PMF07} have distributions that peak at  
lower speeds with tails up to $\simeq 30 c$.  At a factor of 2-3 times lower spatial resolution, 
these studies sample jet emission further from the core region.  So the speed difference may
be due to jets already starting to slow down at these length scales, or perhaps the jet 
components that survive to these distances are systematically different than those closer 
to the core. 

\subsection{Acceleration and Collimation}

Some models of jet formation and collimation have jets being fully accelerated and collimated
very near the super-massive black hole/accretion disk system on length scales much smaller 
than those probed by VLBI\cite{S05} while other models extend this process of much larger length
scales\cite{VK04} where VLBI observations may be able to see acceleration in action.  It is also 
unknown where jets begin to slow down before they reach kiloparsec scales. Very long
time baseline VLBI studies have the opportunity to address these issues by tracking individual
jet components over a very large number of epochs.  We can study acceleration by looking for
changes in the apparent velocity vector $\vec{\beta}_{obs}$ as described in Ref.~\refcite{HKK09}:

\begin{equation}
\frac{d\beta_{\parallel obs}}{dt_{obs}} = \frac{\dot{\beta}\sin\theta+\beta\dot{\theta}(\cos\theta-\beta)}{(1-\beta\cos\theta)^3
}
\end{equation}

\begin{equation}
\frac{d\beta_{\perp obs}}{dt_{obs}} =
\frac{\beta\dot{\phi}\sin\theta}{(1-\beta\cos\theta)^2},
\end{equation}

\vspace{0.15in}

\noindent which are the parallel and perpendicular components of the observed acceleration on the sky.  
$\dot{\beta}$, $\dot{\theta}$, and $\dot{\phi}$ are the intrinsic rates of change of the component speed, angle to
the line of sight, and azimuthal angle respectively.  

If observed parallel accelerations are due entirely to changes in the component's 
Lorentz factor, $\Gamma$, there is a simple relation with the 
observed quantities\cite{HKK09}:

\begin{equation}
\frac{\dot{\Gamma}}{\Gamma} = \frac{\dot{\beta}_{\parallel obs}}{\beta_{obs}} \frac{\beta^2}{\delta^2}
\end{equation}

\vspace{0.15in}

\noindent where the Doppler factor $\delta = 1/(\Gamma(1-\beta\cos\theta))$.  If $\delta$ can
be estimated, this approach allows one to measure the ratio $\dot{\Gamma}/\Gamma$ which 
can be compared to physical models for acceleration or deceleration.

A key question is the extent to which apparent accelerations in jet motions can be assigned
to changes in the Lorentz factor of the jet feature or changes in the angle to the line of 
sight.  In their kinematic analysis of the 2cm Survey, Ref.~\refcite{K04} found that 30\% of the jet components they
studied for non-radial motion were indeed moving on a vector mis-aligned with the radio core,
indicating that these components had changed their trajectory since being ejected from the base
of their jet. In general these ``non-ballistic'' jet features were misaligned toward the direction
of the next structure in the jet, indicating that they were following pre-established channels\cite{K04}.  
These
results were confirmed by both the RRFID kinematic analysis\cite{PMF07} and the MOJAVE kinematic analysis,
which extended the time baseline of the 2cm Survey by several years\cite{LCH09}.  When only the most well 
determined MOJAVE jet components were analyzed (those suitable for acceleration analysis) the fraction
with non-ballistic motion increased to nearly 50\%\cite{HKK09}.

The long time baseline and very large number of epochs in the MOJAVE program allowed Ref.~\refcite{HKK09}
to extract
the 203 best best studied jet features in their sample for direct acceleration analysis.  
They measured the
apparent parallel and perpendicular accelerations.  Parallel accelerations are along the 
component motion, indicating changes in apparent speed, and perpendicular accelerations indicate
changes in direction.  By studying the ratio of these two quantities, they concluded
that intrinsic changes in the Lorentz factor of jet components were common\cite{HKK09}.  They observed a 
tendency for jet components with increasing apparent speed to be closer to the base of their 
jets than components with decreasing apparent speed, suggesting that the jet flow may increase 
in speed near the base of the jet and decrease further out\cite{HKK09}, although this assumes
the observed pattern changes are reflective of the underlying flow. Ref.~\refcite{J05} also observed a
tendency for positive acceleration of apparent speeds near the base of jets in their 
sample of 15 blazars at 43 GHz. While the authors interpreted these results as evidence that the
jets were bending away from the line of sight (and closer to the optimum angle for superluminal
motion)\cite{J05}, the MOJAVE program results described above suggest that these changes may 
be better explained (on average) by increases in the Lorentz factors of those components.  

\subsection{Jet Opening Angles and Morphology}

Ref~\refcite{PKLS09} used the most recent MOJAVE epochs available at the time to study the
correlation between apparent jet opening angle and Gamma-ray brightness of AGN jets.  They
found that Gamma-ray blazars had significantly larger apparent opening angles
than non-detected AGN, indicating that these Gamma-ray bright AGN are more closely aligned
with the line of sight and therefore more highly beamed\cite{PKLS09}.  They also combined
their apparent opening angle results with Doppler factor measurements from Ref.~\refcite{Hovatta09}
and apparent speeds from the MOJAVE program\cite{LCH09} to extract intrinsic opening angles, finding
average jet opening angles on parsec scales of $1.2\pm0.1$ degrees for quasars and $2.4\pm0.6$ 
degrees for BL Lacs.

It has been well known for some time that some of the best studied blazar jets show ejections 
of new jet features at multiple position angles\cite{AC98}\cdash\cite{SCS03}.  The very 
long time baselines in the MOJAVE program have allowed us to sample a wide range of component
ejection and propagation behavior in jets.  For jets with
a range of ejections angles, the structure at any single epoch may only show those areas
of the jet that have been recently illuminated by a passing component.  This can give the false
impression of bent or twisted jet trajectories when the component motion is actually largely 
ballistic.  An example is shown in figure 2, where a single epoch image of the quasar 1308$+$326
is compared to a 'stack' of 58 epochs from the MOJAVE program\cite{LAA09}.  The single epoch
image gives the impression of a sharply bent trajectory, whereas the reality is that the features
are moving outward\footnote{Although note that component 5 in figure 2 is non-ballistic, indicating
that it has changed its trajectory since ejection to be about 8 degrees to the south, relative 
to a ballistic trajectory from the base of the jet.} in a broad cone on the sky which is revealed in 
the stacked image. Images of this kind suggest a possible new way to study jet opening angles 
and collimation through time-averaged morphology over the course of many outbursts.

\begin{figure}[pb]
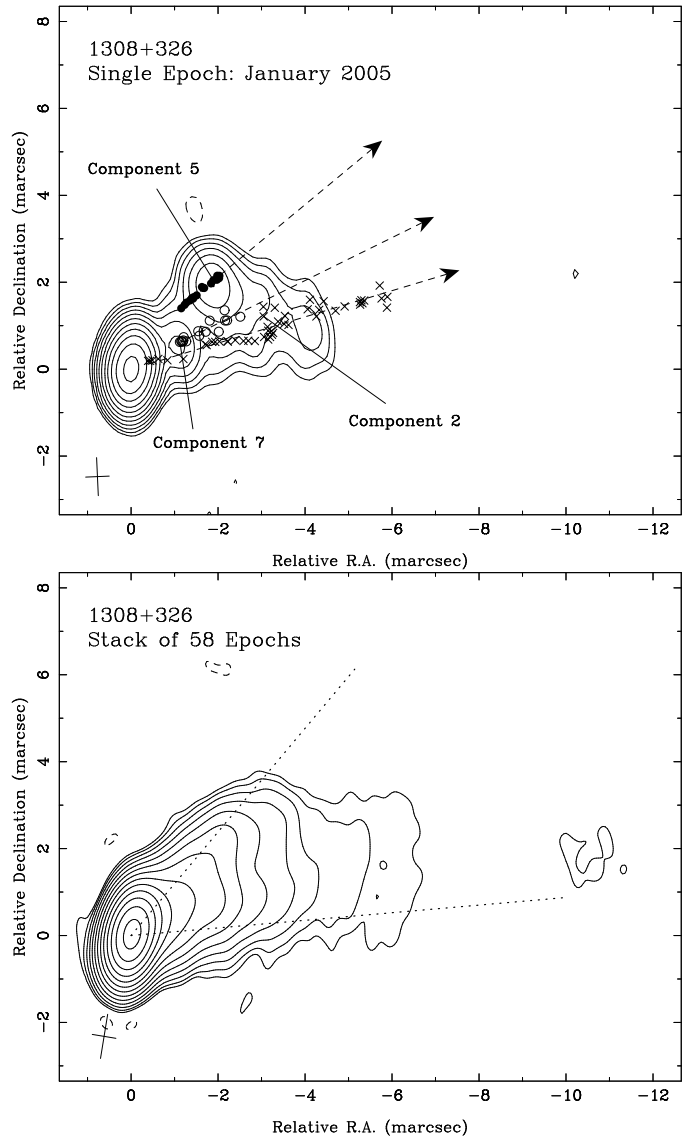

\centerline{\psfig{file=Homan_1_f2a.eps,width=7.5cm,angle=-90}}
\centerline{\psfig{file=Homan_1_f2b.eps,width=7.5cm,angle=-90}}
\vspace*{8pt}
\caption{15 GHz VLBA Images of the quasar 1308$+$326 at $z=0.997$, both in a single MOJAVE epoch, top panel,
and stacked over 58 MOJAVE and archival epochs processed by the MOJAVE program, bottom panel 
(http://www.physics.purdue.edu/astro/MOJAVE/).
The apparent motion of three components are over-plotted on the top panel to illustrate the 
outward nature of the motions.  Components 2 and 7 are consistent with ballistic 
trajectories while component 5 is non-ballistic by about 8 degrees, indicating 
that it has changed its motion since being ejected from the core. The bottom panel
has dotted lines indicating the approximate opening cone revealed in the stacked
image.\label{f2}}
\end{figure}

\section{Conclusions}

Very Long Baseline Interferometry studies of parsec-scale jets with structural and polarization
information gathered over very long {\em time} baselines are producing interesting new 
results. By observing multiple jet features over long periods of time, we are able to obtain
better estimates of maximum apparent jet speeds in individual jets and use these when studying
the Lorentz factor distribution of AGN jets as a whole.  Lorentz factors up to 10-20
appear to be relatively common in blazar jets with a tail extending up to a maximum Lorentz factor 
of $\sim 50$ for the blazar population\cite{J05,LCH09}.  When individual jet features are 
tracked for a large number of epochs it becomes possible to study their apparent acceleration 
both due to bending and due to intrinsic changes in Lorentz factor.  It appears that real changes in 
Lorentz factors for jet features play an important role in observed accelerations\cite{HKK09}.  
Whether these accelerations reflect changes in the underlying flow is unclear; however a tendency for jet 
features with increasing speed to appear closer to the base of their jets than features
with decreasing speed may be evidence of corresponding changes in the flow speed on these
length scales\cite{HKK09}. Finally, the ejection of multiple components along differing position
angles in jets, when studied over very many epochs, allows the production of stacked, time-averaged
images of the jet morphology\cite{LAA09}, suggesting a possible new way of studying jet opening angles 
and collimation.

\newpage

\section*{Acknowledgments}

I would like to thank all the members of the MOJAVE program 
including Matt Lister, Yuri Kovalev, Ken Kellermann,
Hugh Aller, Margo Aller, Tigran Arshakian, Andrei Lobanov, Tuomas 
Savolainen, Anton Zensus, Eduardo Ros, Matthias Kadler, Neil Gehrels,
Julie McEnery, Marshall Cohen, Alexander Pushkarev, and Talvikki
Hovatta.  The MOJAVE program is funded at Purdue University through
National Science Foundation grant AST-0807860 and NASA-Fermi 
grant NNX08AV67G. D. Homan was funded by National Science Foundation
grant AST-0707693.

\def\aj{{\it The Astronomical Journal}} 
\def\apj{{\it The Astrophysical Journal}}
\def\apjl{{\it The Astrophysical Journal, Letters}}
\def\apjs{{\it The Astrophysical Journal, Supplement}}
\def\apss{{\it Astrophysics and Space Science}}        
\def\aap{{\it Astronomy and Astrophysics}}
\def\aaps{{\it Astronomy and Astrophysics, Supplement}}
\def\mnras{{\it Monthly Notices of the Royal Astronomical Society}}


\end{document}